\documentclass[12pt]{article}
\usepackage{graphicx}
\usepackage{subfig}
\usepackage{amssymb,amsmath,amsfonts,palatino,amsthm}
\usepackage{amssymb}
\usepackage{epstopdf}
\newcommand{\f}{\begin{equation}}
\newcommand{\ff}{\end{equation}}

\begin{document}

\small
\title{The thermodynamics of quantum spacetime histories}
\author{Lee Smolin\thanks{lsmolin@perimeterinstitute.ca} 
\\
\\
Perimeter Institute for Theoretical Physics,\\
31 Caroline Street North, Waterloo, Ontario N2J 2Y5, Canada\\
and\\
Department of Physics and Astronomy, University of Waterloo\\
and\\
Department of Philosophy, University of Toronto}
\date{\today}
\maketitle

\begin{abstract}

We show that the simplicity constraints, which define the dynamics of spin foam models, imply, and are implied by, 
the first law of thermodynamics, when the latter is applied to 
causal diamonds in the quantum spacetime.  
This result reveals an intimate connection between the holographic nature of gravity, as reflected by the Bekenstein entropy,  and the fact that general relativity and other gravitational theories can be understood as constrained topological field theories.

To state and derive this correspondence we describe causal diamonds in the causal structure of spin foam histories and generalize arguments 
given for the near horizon region of black holes by Frodden, Gosh and Perez\cite{FGP} and Bianchi\cite{Eugenio1}.  
This allows us to apply a recent 
argument of Jacobson\cite{Ted2015} to show that if a spin foam history has a semiclassical limit described in  terms of a smooth metric geometry, that geometry satisfies the Einstein equations.  

These results suggest also a proposal for a quantum equivalence principle.

This paper is dedicated to the memory of Jacob Bekenstein.

\end{abstract}

\newpage

\tableofcontents

\normalsize
\newpage

\section{Introduction}

Ever since Jacob Bekenstein proposed that black holes have entropy\cite{jb}, the idea that the dynamics of spacetime are expressions of the laws of thermodynamics has been contemplated.  Indeed, Bekenstein was inspired by the observations of Bardeen, Carter and  Hawking that black holes obey a set of laws analogous to those of thermodynamics\cite{bhmech}.  Bekenstein's groundbreaking 1972 paper was followed by Hawking's discovery of black hole radiation\cite{Hawking-bh}, as well as the work of  Davies\cite{Davies}, Fulling\cite{Fulling} and Unruh\cite{Unruh} on Unruh radiation.  Reflecting on these results, Candelas and Sciama\cite{cs} and others posed the question of whether the Einstein equations were a result of the statistical mechanics of some quantum gravity theory.  

Ted Jacobson took a large step towards this goal in his 1995 paper in which he showed that the Einstein equations emerge as the equation of state of some atomic structure of spacetime\cite{Ted95}\footnote{See also \cite{Paddy}.}.  

There are two ways to read these result connecting 	gravity, the quantum and  thermodynamics.  The conservative point of view is that the laws of thermodynamics and the Einstein equations both emerge at the semiclassical level.  Here I provide evidence for a deeper possibility, which is that the first law of thermodynamics is expressed by  the microscopic dynamics of the quantum spacetime.  

To see how this arises we can trace the story of responses to Bekenstein's great discovery.

First, in 1993, Louis Crane proposed that quantum gravity be closely related to topological quantum field theory, as the latter is characterized by Hilbert spaces of states appearing on boundaries, as is suggested by the Bekenstein bound\cite{Louis1}.  
Indeed,  Plebanski\cite{Plebanski}, Capovilla, Dell and Jacobson\cite{CDJ}, and others had shown that general relativity is elegantly expressed as a topological field theory, called $BF$ theory, modified by the imposition of certain constraints.  This is the theory of a two form-the $B$ field, interacting with a gauge field.  The constraints are known as classical simplicity constraints, as they require that that $B$ field be simple.  More precisely, the Einstein's equations are a consequence of constraining the gauge degrees of freedom of the $B$ field.  These constraints  reduce the gauge invariance of the theory.  This in turn  liberates certain bulk gauge modes to become physical, and they are exactly the massless spin two modes.  At the same time, the boundary degrees of freedom remain those of the topological field theory.   

Moreover, as shown in \cite{linking}, when boundary conditions are imposed within this framework which code the presence of an horizon, the boundary dynamics is exactly 
Chern-Simons theory and the boundary Hilbert space has a finite dimension that grows with the 
exponential of the area, as defined in loop quantum gravity\footnote{This insight has been recently deepened in \cite{aldoetal}.}.  

This led to an understanding of black hole entropy in LQG\cite{Kirill-bh,isolated}, but left the relation between area and entropy dependent on a free parameter: the Immirzi parameter, which gives the area gap.    This ambiguity was resolved when the original canonical picture of isolated horizons on which Chern-Simons theory is induced was developed in a spacetime or spin foam language by Frodden, Gosh and Perez\cite{FGP} and Bianchi\cite{Eugenio1}.  The dependence on the Immirzi parameter was shown to be a feature of the ensemble of states at fixed area, and is lifted when we consider instead ensembles based on constraining the temperature or energy.  For these ensembles, the entropy is exactly the Bekenstein-Hawking entropy, with the correct $\frac{1}{4}$.   In particular, the ensembles of fixed boost energy, seen by a stationary near horizon observer, plays a key role in \cite{FGP, Eugenio1}, as it does in this paper\footnote{For a review of different approaches to black hole entropy within LQG, see \cite{AF}.}. 

The key to these spin foam derivations of the Bekenstein entropy is the imposition of a quantum version of the simplicity constraint, in the measure of the path integral of the corresponding topological quantum field theory.  This again liberates the spin two modes. These  simplicity constraints\cite{FK,EPRL}, defined by (\ref{simp}) below, reduce the partition function of a topological field theory to the partition function of general relativity. 

Meanwhile, around the same time as Crane's papers,  't Hooft, also inspired by Bekenstein's discovery, proposed the holographic principle\cite{tHooft-holo}.  This was quickly taken up by Susskind in the context of string theory\cite{Lenny-holo}.  This inspired Maldacena to propose the AdS/CFT correspondence\cite{Malda1}.  This has become a cornerstone of contemporary physics, with many examples developed in string theory.  In addition, there are indications that this reflects a deep correspondence between conformal field theories and diffeomorphism invariant theories that transcend any single realization\cite{AdSCFT-shape}.

By making use of the AdS/CFT correspondence, the authors of several ingenious papers were recently able to show that the linearized Einstein equations are a consequence of entanglement\cite{AdS-grav-entanglement}.  These suggest a principle of maximal entanglement which appears to generalize the fact that the vacuum of a QFT on Minkowski spacetime is maximally entangled.  

This very recently inspired Jacobson to return to the subject and rework his 1995 argument as a demonstration that the Einstein equations express such a principle of extremal entanglement\cite{Ted2015}.   

Like his 1995 paper, Jacobson's derivation can be understood as a schema for deriving the Einstein equations as the statistical thermodynamics of a discrete theory of quantum spacetime.  In this paper we realize this schema in the context of spin 
 foam models.

 The key result we find is a close relationship between thermodynamics and the simplicity constraint of spin foam models\cite{FK,EPRL,reviewSF}. Our major result is that under certain conditions, {\it the simplicity constraint implies directly the first law of thermodynamics.  }               We also show that under reasonable assumptions the first law implies the simplicity constraints.
 These results extend and generalize previous results\cite{FGP, Eugenio1,carloetal}.
 
 This result ties together the holographic principle with the understanding of general relativity as a constrained topological field theory.

\begin{figure}[t!]
\centering
\includegraphics[width= 0.5\textwidth]{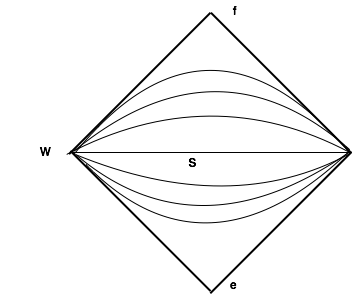}
\caption{A causal diamond defined by two events, $f$ and $e$.  $W$ is the corner, or waist, bounding a space like three-disk, $S$.}
\label{causaldi1} 
\end{figure}

There are actually several closely related results, which are the following.

\begin{enumerate}

\item{}The first result is not new, it is rather an interpretation of a result given in different forms by  Carlip and Teitelboim\cite{CT}, 
Massar and Parentani\cite{MP,JP},  Bianchi and Wieland\cite{BW}, Frodden, Gosh and Perez\cite{FGP} and the author\cite{ls-bh}.  They consider the exterior of a black hole horizon or, more generally, a causal diamond.  These regions have a bifurcation two sphere, $W$ or, more generally, a ``corner" or  ``waist", which is fixed under a family of boost transformations. (See Figure 1.) The simplest example of this are the transformations generated by boost killing fields in Rindler spacetime, but these exist for any causal diamond.  These authors compute the contribution to the Hamiltonian which generates such a boost, coming from a boundary term at the corner.  (There are also bulk terms, but they vanish on solutions because they are proportional to constraints.) They show that
\f
H(W)_{Boost} = \frac{1}{8 \pi G} A(W)
\label{boost1}
\ff
where $A(W)$ is the area of $W$, computed by the two metric induced on $W$.
Now, assume that $W$ is indeed the bifurcation two sphere of a stationary black hole.  Then $A(W)$ is the area of the horizon and we can make use of the classical second law of black hole mechanics which tells us that under a physical process that converts one stationary black hole into another, $A(W)$ can never decrease\cite{bhmech}.  This suggests we define an entropy by introducing a discrete
unit of area, $\Delta a$, so that\footnote{As \cite{carloetal} emphasize, even if (\ref{firstc1}) is a classical result, the fact that we have to introduce a finite unit of area $\Delta a$ to give the surface entropy a meaning points to a hidden role of quantum spacetime in defining the thermodynamics of the gravitational field.}
\f
S(W) = \frac{A(W)}{\Delta a}
\ff
Then our relation for the boost hamiltonian, (\ref{boost1}) must be related to the first law of black hole mechanics.  
We do not show this here, but we note that it can be written in the suggestive form,
\f
H(W)_{Boost} = T_B S(W)
\label{firstc1}
\ff
where the universal, Unruh-like, boost temperature is $T_B= \frac{\Delta a}{8 \pi G}$.  Of course when we pick 
$\Delta a = 4 G\hbar$ we get the usual results, but the point is that (\ref{firstc1}) is a classical relationship.
This relation (\ref{firstc1}) is more general than the first law of black hole mechanics, because it holds on the corner of every 
causal diamond.
We can call (\ref{firstc1}) the {\it first law of classical spacetime dynamics}.  FGP call it the local form of the first law\cite{FGP}.   Indeed, as Jacobson shows\cite{Ted95,Ted2015}, this quasi thermodynamic relation implies the Einstein equations.  

With this classical prelude we go on to see that it has a consequence for quantum gravity, which is the simplicity constraint.

\item{}The relation between (\ref{firstc1}) and the simplicity constraint emerges as soon as we express the operators for boosts
and area in the language of loop quantum gravity.  For experts, the result can be telegraphed in a few lines.

In a spin foam model a space like two surface like $W$ is represented by a set of triangles, $W= \cup \triangle$, in the simplicial decomposition of the spacetime.  Each triangle is the home of a representation of the lorentz group, ${\cal S}_\triangle$.
Each $\triangle$ is also oriented as part of the boundary of a tetrahedron by a space like unit vector, $n_a^\triangle$.
The boost Hamiltonian is equal to a boundary term 
\f
H(W)_B = \hbar \sum_{\triangle \in W} \hat{K}^a n_a^\triangle
\label{boost12}
\ff
plus bulk terms which are linear combinations of quantum constraints. $ \hat{K}^a$ is the generator of boosts 
in ${\cal S}_\triangle$. The area operator is 
\f
\hat{A}_W = 8 \pi \hbar G \gamma \sum_{\triangle \in W} \hat{L}^a n_a^\triangle
\label{area1}
\ff
where $ \hat{L}^a$ is the generator of rotations in ${\cal S}_\triangle$ and $\gamma$ is the Immirzi parameter.

We should now consider that any triangle can be part of a corner of some boost transformation.  Moreover in these case the normals will vary.   Hence, 
these three relations, (\ref{firstc1}, \ref{boost12},\ref{area1}) together imply that when $\Delta a = 4 \hbar G$, physical states satisfy a constraint separately on each triangle, independent of the normals,
\f
< \left ( \hat{K}^a - \gamma \hat{L}^a \right )   > =0
\label{simp0}
\ff
But these are the simplicity constraints that define the spin foam models\cite{FK,EPRL,reviewSF,Eugenio1}.  Hence, the simplicity constraints are a consequence of (\ref{firstc1}),  the first law of classical spacetimes.  They express this classical relation on quantum states.  

The other  results require a bit more structure to describe.  

\item{}Given a spin foam history, $X$, consider a closed space like two surface made of triangles, $W= \sum_\triangle$ bounding a three disk, $\Sigma$.  To $W$ there is associated a Hilbert space, 
${\cal H}(W)= \otimes_\triangle {\cal V}_\triangle \bigotimes {\cal H}_\Sigma$.  Here ${\cal V}_\triangle$ is an infinite dimensional reducible representation of $SL(2,C)$ and ${\cal H}_\Sigma$ contains bulk states that depend on degrees of freedom in the interior of $\Sigma$.  Let $H_B (W)$ be a  corresponding generalized boost (or bubble) Hamiltonian, to be described below. A generalized boost is a transformation that evolves the interior of $\Sigma$ forward in time while leaving its boundary and exterior fixed.  Let $S(W)$ be the entanglement entropy which is a consequence of tracing a global state over degrees of freedom in the exterior.  Then we show that the simplicity constraints, (\ref{simp0}),
imply
\f
< H_B (W) > = T_U S(W)
\label{boost2}
\ff
where
\f
T_U= \frac{\hbar}{2 \pi}
\label{TU}
\ff
is the (angular) Unruh temperature.

We can  call (\ref{boost2}) the {\it first law of quantum spacetime dynamics.}  Thus, we show that the classical relation (\ref{boost1}) implies the quantum constraint,
(\ref{simp0}) and that (\ref{simp0}),  in turn, implies a quantum form of (\ref{boost1}), which is (\ref{boost2}).

(\ref{boost1}) and (\ref{boost2}) have the form of the first law, but they are not yet that law.  To invoke it we need still more structure.  One might think one has to introduce black holes, but it turns out that we can work in a more general context which is  causal diamonds\cite{Ted2015}.  

\item{}So now, let $W$ be the two surface bounding the ``waist" of a causal diamond (all defined for a class of spin foam models below).  Let 
$\delta Q$ be the expectation value of matter energy density crossing $\Sigma$, while $\delta S(W)$ is the change in entropy from a casual diamond of the same spatial volume (of an extremal slice  $\Sigma$) in a flat simplicial spacetime.  Then we show that in a semiclassical approximation,  the usual first law of thermodynamics holds
\f
\delta Q = T_U \delta S(W)
\ff

\item{}In a certain semiclassical regime, to be defined below, in a weak sense, 
we can follow Jacobson's 2015 derivation to recover the Einstein equations\cite{Ted2015}.  It should be emphasized that the result is weak in two senses.  First, we do not show that a spin foam model has a good semiclassical limit.  We show rather that {\bf if} it does, and {\bf if} that limit is described in terms of a slowly varying metric geometry, {\bf then} that metric satisfies the semiclassical Einstein equations\footnote{ We note that this weak recovery of general relativity from spin foam models has been shown previously in other ways, including a large spin or semiclassical limit\cite{semisf} as well as by mimicing the logic of Jacobson's 1995 paper\cite{LS-thermobh1}. }.  Second, the expectation value of the energy momentum tensor that appears in those Einstein equations is defined from the thermodynamics, as the source of the heat flow, and not from a quantization of microscopic degrees of freedom.

\end{enumerate}

 Before  going on to the details, we make some comments on the results\footnote{After this paper was in draft, Aldo Riello  pointed out  \cite{carloetal} where some of the same results are noted, but in the course of making a very different, though  complementary,  argument.}.
 
 It is interesting to note that the Unruh temperature appears in this derivation, as the temperature of a subsystem defined by a causal domain.   This suggests that when a generalized boost creates an horizon, tracing the vacuum state of the quantum gravity theory by degrees of freedom outside that horizon thermalizes the state.  If correct, this reflects a maximization of information shared between the regions of space on each side of the horizon, as in the vacuum of Minkowski spacetime.

One can also derive the Unruh temperature for the boosted frames in a causal diamond, following the calculation of 
Bianchi\cite{Eugenio1}.  In this case the Bekenstein-Hawking entropy, with the correct $\frac{1}{4}$, follows as a consequence of the 
simplicity constraint of the spin foam model.  This, indeed, was essentially the result of Bianchi, for quantum Rindler domains representing the near horizon quantum geometry.  What we show here is that the logic of Bianchi's derivation applies more generally to causal diamonds of spin foam models, and that at the root of these results is a very general connection between
the first law and the simplicity constraint.

Finally, these results suggest a form of the quantum equivalence principle, which has been long sought\cite{cs}.  In flat spacetime an accelerating observer sees a region of spacetime limited by an horizon, which is generated by a two surface, $W$, fixed by a boost.  We can generalize the notion of a boost to mean any evolution of a region of quantum or classical spacetime that fixes a two surface, $W$.  This two-surface $W$ divides a spacial slice of the universe into two parts.  In flat spacetime these are maximally entangled with each other.  We can posit that this is also true in a dynamical quantum spacetime. As this generalizes a property of flat spacetime it is appropriate to call it a version of the equivalence principle.  

The result is that an observer inside the  causal diamond sees a thermal state, given by
\f
\rho_W = e^{-H_{boost}(W)/T_U}
\ff
 where $H_{boost}(W)$ is the generator of the boosts that fix $W$.   As will be explained, because of 
 refoliation invariance in the interior of the causal diamond, this is unique.

 In Jacobson's 2015 paper a key role is played by causal diamonds of a classical spacetime.  In the present paper we work with an analogous notion defined using the causal structure of a spin foam model.  Our first job is then to review spin foam models, and the discrete causal structures they carry.  These causal structures are induced when  the spin foam history is constructed by sequences of dual Pachner moves acting on an initial state, as was shown in \cite{Fotini1} by Markopoulou.   We find that we are able to describe causal domains and their boundaries in sufficient detail to be useful for defining physical observables.  
 
Having control of these causal structures will allow us to define and study causal diamonds in spin foam histories.  We give a general description of causal diamonds in these discrete causal structures,  this allows us to define observables for spin foam models in terms of expectation values of currents defined on the boundaries of causal diamonds.  
 
 The thermodynamics of these quantum spacetimes is then defined by vacuum expectation values on the spatial boundary of the causal diamond, where by  vacuum we mean the case that the currents on the null boundaries vanish.
 
Physical quantities relevant to the low energy or semiclassical limit will, as Dittrich has emphasized, have to be defined following a process of coarse graining and renormalization. Also, as she emphasizes, physical observables are to be defined by coarse graining boundaries\cite{Bianca1}.    It is then important to show that the connection between the simplicity constraints and the first law emerges in a way that is largely independent of these processes and the details of how they are carried out.
To accomplish this we give a very brief sketch of these processes.  The key point is that the linear simplicity constraints, acting on boundary observables, apply just as well to coarse grained and renormalized quantities, because they 
are 
linear.

 
 In the next section we review the construction of causal spacetime histories, which we use in section 3 to describe causal diamonds in those histories.  Section 4 describes the quantumm mechanics of causal diamonds and their associated observables in both the  canonical and path integral language.  We use these results in section  5 to present the main results, which are  a set of relationships between the simplicity constraints and different versions of the first law.  Section 6 is brief and sketches the application of Jacobson's 2105 argument in \cite{Ted2015} to show that {\bf if} a spin foam model has a suitable semiclassical limit, describerle in terms of slowly varying metric and matter fields, those fields obey the Einstein equations.  Some comments are presented in the last section 7.
 
\section{Summary of causal spin foam models}

We start by summarizing the basic structures used to define spin foam models\footnote{For reviews of spin foam models,
see \cite{reviewSF}.}.

\subsection{Causal spin foam models}

\begin{figure}[t!]
\centering
\includegraphics[width= 0.5\textwidth]{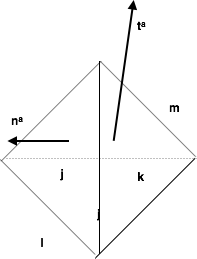}
\caption{A tetrahedron and its labeling}
\label{tetra1} 
\end{figure}

\begin{itemize}

\item{}We define a Hilbert space of states, $\cal H$, arising from the quantization of general relativity, on a spacetime 
manifold with topology $\Sigma \times R$, where $\Sigma$ is a three dimensional manifold which is either compact or compact with a boundary.  In the latter case there will be boundary  conditions imposed on $\partial \Sigma$.  A basis for $\cal H$ is 
given by spin networks embedded in $\Sigma$, modulo diffeomorphisms of $\Sigma$.  A dual description of this basis is triangulations of $\Sigma$, in terms of tetrahedra, representing a space like slice, i i.e. a three dimensional simplicial complex, made by gluing tetrahedra together.  The faces of the tetrahedra are triangles which are labeled by representations $j$ of $SU(2)$.  The tetrahedra themselves are labeled by intertwiners in the product of representations of its triangles.

\item{}A spacetime history is denoted $X$.  A history is a four dimensional simplicial complex that  interpolates between an initial state $|in>$ and a final state
$|out>$.  

\item{}Events of $X$, denoted $e_1,e_2, \ldots$,  are dual to four simplifies $S$.  A four simplex represents a bubble move that evolves a small region of a state, consisting of a small number of tetrahedra. 

\item{}Each history $X$ interpolating between $|in>$ and $|out>$ contributes a complex amplitude ${\cal A} (X)$ to the transition
from $|in>$ to $|out>$.  The total amplitude is the sum,
\f
{\cal A}[ |in> \rightarrow |out> ] = \sum_{X | \partial X = | in > - | out> } {\cal A} (X)
\ff
The amplitudes are computed by the following rules.

\item{}The space-like triangles of the four simplifies that make up a history are labeled by representations of the lorentz group, $(\rho,j)$.  
We associate to each space-like triangle a Hilbert space which is the corresponding representation space, ${\cal V}_{(\rho, j )}$.
Acting on this Hilbert space
are generators of boosts, $\hat{K}^a$ and rotations, $\hat{L}^a$.   Associated to each triangle, $\triangle$, is also a unit normal, $n_a$, which lives in an auxiliary flat spacetime, $M^4$.  

The sum of the unit normals for the triangles of a tetrahedra are constrained to vanish.
\f
\sum_{\tau \in T} n^a_\tau =0
\ff

We will be interested in a class of simple representations, which are given by a map from representations of $SU(2)$.
\f
Y_\gamma : j \rightarrow (\gamma (j+1 ) , j )
\ff
where $\gamma$ is the Immirzi parameter.  These representations satisfy the simplicity constraint,  ${\cal S}$
\f
<\Psi | \hat{\cal S}^a |\Psi > =  <\Psi | \left ( \hat{K}^a - \gamma \hat{L}^a  \right )  |\Psi >=0 
\label{simp}
\ff


\end{itemize}

\subsection{Causal structures in spin foam histories}

We review the causal structure of spin foam models, first proposed in \cite{Fotini1}.  

\begin{itemize}

\item{}A four simplex $S$ contains $5$ tetrahedra, $T$.  $n$ of these are in ${\cal PS}(S)$, the past set of $S$.
$5-n$ of these are in ${\cal FS }(S)$, the future set of $S$.

\item{}Each tetrahedra (except those in the future or past boundary of $X$) is in the future set of one four simplex $S_1$ and the 
past set of another four simplex $S_2$.  We say that $S_2$ is in the immediate future of $S_1$:
\f
S_2 \in {\cal  IF} (S_1 )
\ff
and  $S_1$ is in the immediate past of $S_2$:
\f
S_1 \in {\cal IP(} S_2 )
\ff
Dual to $T$ is a link connecting the event, $e_1$,  dual to $S_1$ to the event, $e_2$, dual to $S_2$.   

We can write also $S_1 = {\cal P}(T)$ , i.e. the four simplex, $S_1$ is the past of the tetrahedron $T$ if the causal link which is the deal of $T$ points directly from the event dual to $S_1$ .  Similarly, $S_2 = {\cal F}(T)$.   

\item{}We assume that the triangles which bound the tetrahedra are all space like.   This will be the case when the four dimensional simplicial complex that defines the spin foam history is constructed from a dual spin network (a union of tetrahedra joined alone triangles that are dual to the edges of a spin network) by a succession of Pachner moves, each representing an event where the dual spin network state is changed locally.

\item{}We say that an event $f$ is in the causal future of an event $e$ if there is a chain of future pointing causal links
taking $e$ to $f$.  We write
\f
f \in {\cal F} (e)
\ff
Similarly we write that $e$ is in the causal past of $f$
\f
e \in {\cal P}(f)
\ff

\end{itemize}

\subsection{Wieland structures on spin foams}

We will also make use of a version of spin foam dynamics introduced by Wieland\cite{WW3+1,WW2+1}.  These make use of causal structures,
based on energetic causal sets\cite{ECS1,ECS2,ECS3}\footnote{The role of causal sets in quantum spacetime was proposed in
\cite{RS-cs} and developed in different ways in \cite{Fotini1,RL,Cohl-cs}.}.  This formulation makes use of a future pointing normal $p_a^T$ associated to each tetrahedron,$T$,  in the auxiliary flat spacetime, $M^4$,  which is constrained two ways.  

\begin{itemize}

\item{}{\bf Conservation.} The five normals of tetrahedra $T$ making up a four simplex, $S$ sum to zero.
\f
{\cal P}_a^S = \sum_{T \in { \cal FS }(S)} p_a^T -\sum_{T \in {\cal PS} (S)} p_a^T =0
\label{Pcon}
\ff
\item{}{\bf Normalization.}Each $p_a^T$ is constrained as if volume were the mass of a relativistic particle.
\f
{\cal C}^T = p_a^T p_b^T \eta^{ab} + V(T)^2 =0
\label{Ccon}
\ff
where $\eta$ is a flat metric, in $M^4$, and the volume of a tetrahedron $V(T)$ is a function of the spins and intertwiners on its faces and bulk.

\end{itemize}

The action for a Wieland spin foam then has a part made of these constraints.
\f
{\cal S}= \sum_S z^a_S {\cal P}_a^S + N_T {\cal C}^T + \ldots
\ff
where $z^a_S$ and $N_T$ are lagrange multilpliers.  Note that the $z^a_S$ live in the dual space to $M^4$, which inherits a flat metric, $\eta_{ab}$ from the metric $\eta^{ab}$ of $M^4$.

\section{Causal diamonds on spin foams}

In this section we use the causal structures we identified in causal spin foams to define causal diamonds and related structures.

\subsection{The boundary of a past set}

We begin with some further definitions.

\begin{itemize}

\item{}Now divide ${\cal P}(f)$ into the disjoint union of two sets
\f
P(f) = {\cal BP} (f) \cup \bar{\cal P}(f)
\ff
called the bulk and boundary of the causal past of $f$.

$g \in {\cal BP} (f)$ if $IF(g) \in {\cal P} (f)$ i.e. if the immediate future of $g$ consists entirely of members of the causal past of $f$.

Otherwise, $g \in \bar{\cal P}(f) $, i.e. $g$ is an event in the boundary of the causal past of $f$.  $g$ is dual to a four simplex, which c an also be considered as residing in the boundary of the past of $f$.

It is useful to extend the notion of the boundary of a past set to tetrahedra and triangles.  These give the three-boundary, 
$\bar{\cal P}(f)_{(3)}$ and two-boundary, 
$\bar{\cal P}(f)_{(2)}$.

\item{}The tetrahedron, $T$ is in $\bar{\cal P}(f)_{(3)}$ if both its past and future four simplifies are dual to events in $\bar{\cal P}(f)$.  

\item{}The triangle $\triangle $ is in $\bar{\cal P}(f)_{(2)}$  if it is in the boundary both of tetrahedra in $\bar{\cal P}(f)_{(3)}$  and tetrahedra in the exterior of ${\cal P}(f)$.  
\end{itemize}

Similar definitions hold for the boundary of a future set.

\subsection{The causal diamond}

\begin{itemize}

\item{}Now we define the causal diamond of two events $e \subset f$.
\f
{\cal C D} (f,e) = {\cal P}(f) \cap {\cal F}(e)  
\ff

\item{}We define the 3-waist of ${\cal C D}(f,e)$,  which we label ${\cal W}(f,e)$ to be the set of tetrahedra $T$ such that $T$ is both in 
$\bar{\cal P}(f)_{(3)}$ and $\bar{\cal F}(e)_{(3)}$.   

\item{}Define the two-boundary of ${\cal W}(f,e)$ to be those triangles of $T \in {\cal W}(f,e)$ which are dual to edges that connect a vertex dual to a tetrahedra in ${\cal W}(f,e)$ to a tetrahedra not in ${\cal CD}(f,e)$.   Denote these triangles by
$\bar{\cal W}(f,e)_{(2)}$.  This is called the 2-waist of the causal diamond.  


\item{}Define the area of the causal diamond, ${\cal A}(f,e)$ to be the sum of the areas of the triangles in $\bar{\cal W}(f,e)_{(2)}$.

\item{}Define a cross=section of ${\cal CD}(f,e)$ to be a connected three surface, anti chain (i.e. mutually acausal) $\Sigma$,  made of tetrahedra in ${\cal CD}(f,e)$ whose boundary is $\bar{\cal W}(f,e)$.  We use the volume operator to define the volume of such an anti chain.  

\item{}Define the maximal cross section $\Sigma (f,e)$ of ${\cal CD}(f,e)$ to be the cross-section with the maximal volume.

\item{}Define the volume of ${\cal CD}(f,e)$ to be the volume of its maximal cross section.  Denote it ${\cal V}(f,e)$.  

\end{itemize}

\subsection{The three-boundary of a causal diamond}

\begin{figure}[h!]
\begin{center}
\includegraphics[width=.5 \textwidth]{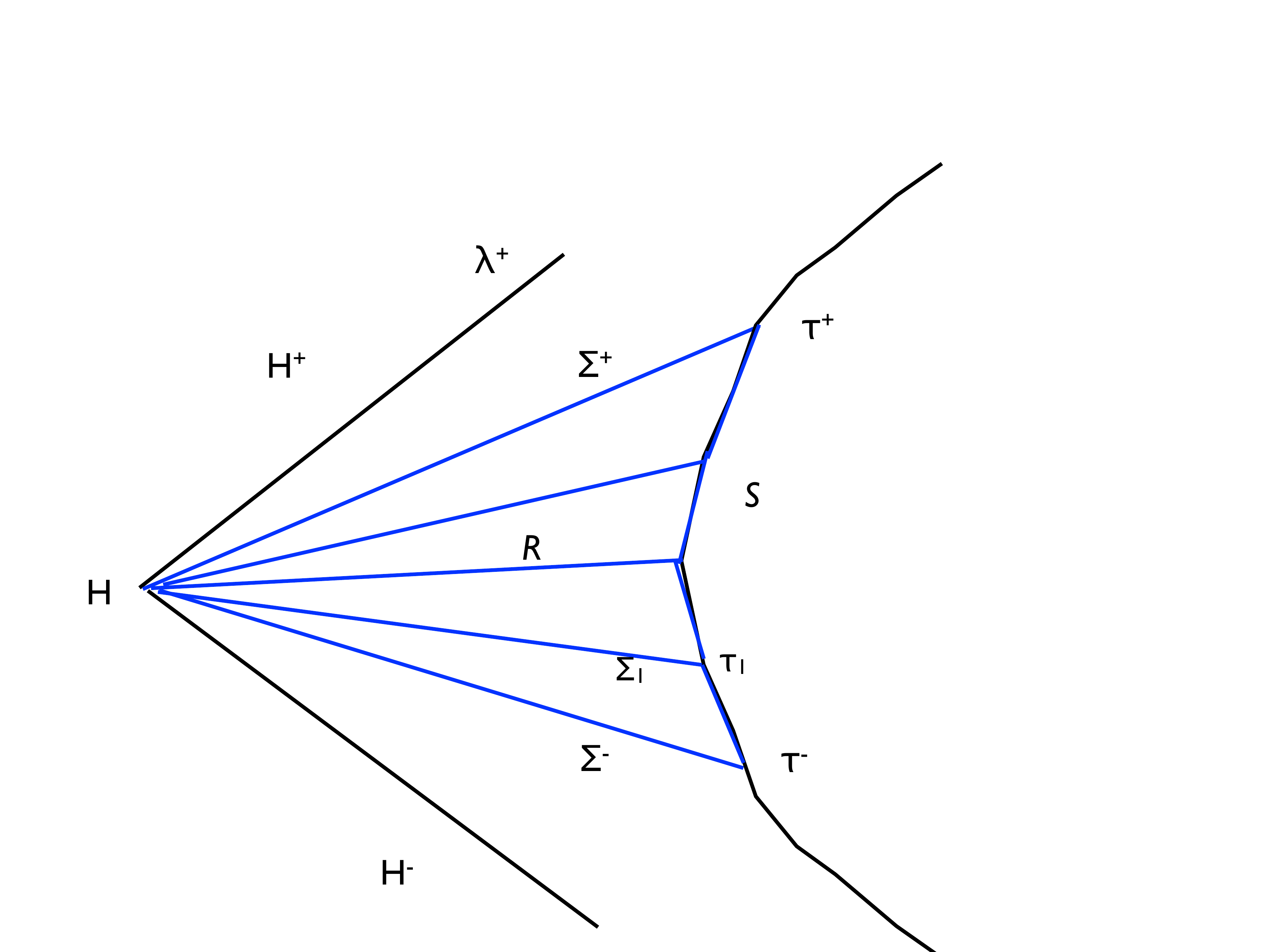}
\end{center}
\caption{The quantum near horizon region $\cal R$ constructed in a spin foam by a series of generalized boosts, here generated by $1 \rightarrow 4$ moves, which are here illustrated in this two dimensional figure by $1 \rightarrow 2$ moves.}
\end{figure}

\begin{figure}[h!]%
    \centering
    \subfloat[A series of $1 \rightarrow 2$ moves. ]{{\includegraphics[width=7cm]{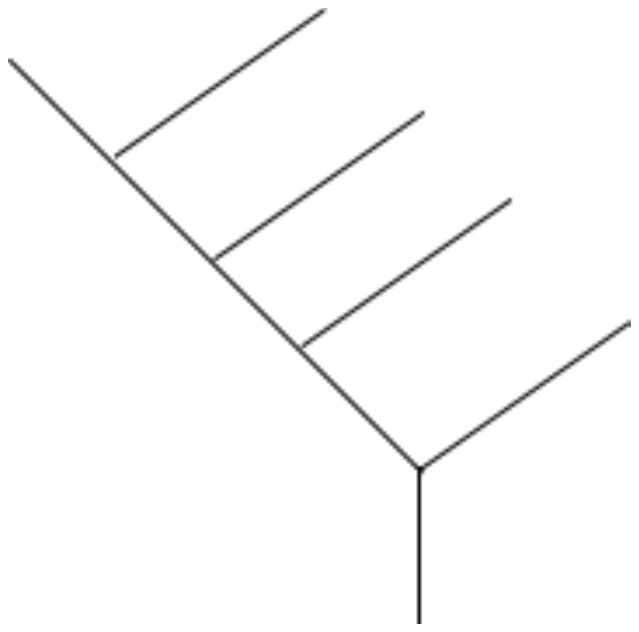} }}%
    \qquad
    \subfloat[In blue we see the dual $1 \rightarrow 2$ moves, which fix a dual vertex, $W$, which is the dimensional reduction of fixing  a triangle in a $1 \rightarrow 4$ move.]{{\includegraphics[width=7cm]{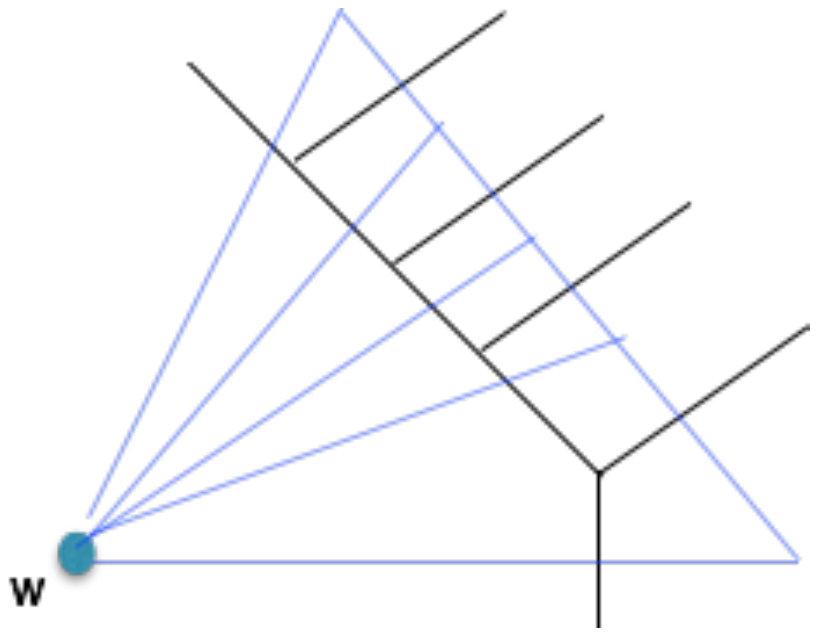} }}%
    \caption{A generalized boost realized in $1+1$ dimensional dual Pachner moves.}%
    \label{fig:example}%
\end{figure}

\begin{figure}[h!]
\begin{center}
\includegraphics[width=.8 \textwidth]{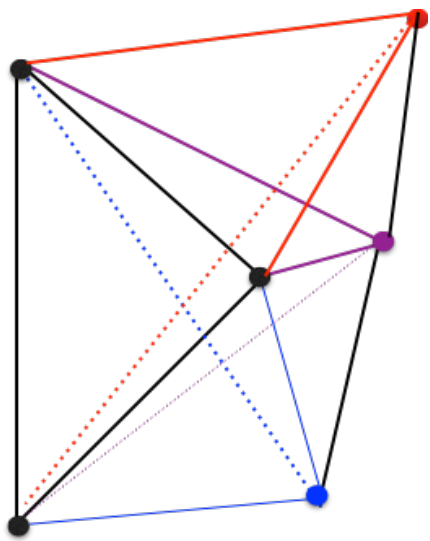}
\end{center}
\caption{Two  $3+1$ dimensional boost moves carried out in succession.  Each fixes the black triangle which common to three tetrahedra, in blue, purple and red.  Two $1 \rightarrow 4$ moves are carried out in succession, which evolve from the blue to the purple to the red tetrahedra.}
\end{figure}

\begin{figure}[h!]
\begin{center}
\includegraphics[width=.8 \textwidth]{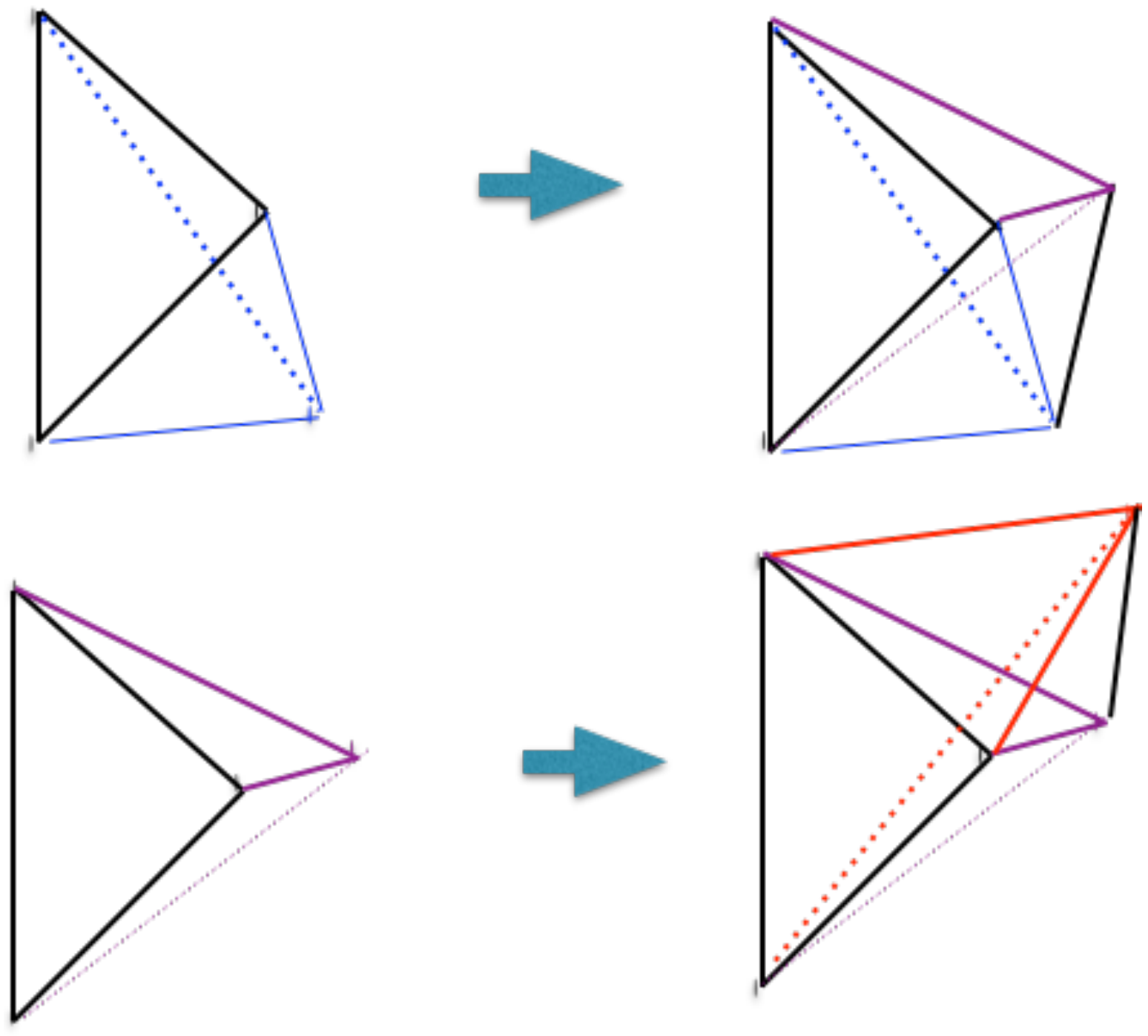}
\end{center}
\caption{Two  $3+1$ dimensional boost moves in the last figure.  Each is a  $1 \rightarrow 4$ move represented by a four simplex that fixes the black triangle.}
\end{figure}

\begin{itemize}

\item{}Define the future boundary of a causal diamond, ${\cal CD}(f,e)$ to consist of the  tetrahedra which are in the boundary of 
${\cal P}(f)$ and also in ${\cal CD}(f,e)$, and denote this set ${\cal I}^+ (f,e)$.

Define the past boundary of a causal diamond similarly, as 
\f
{\cal I}^- (f,e) = \bar{\cal F}(e) \cap {\cal CD}(f,e)
\ff

\item{}Define the $3-$ boundary of a causal diamond to consist of 
\f
\bar{\cal CD}(f,e)_{(3)} = {\cal I}^+ (f,e) \cup {\cal I}^- (f,e) \cup {\cal W}(f,e)
\ff

\end{itemize}

\subsection{The two-boundary of a causal diamond}

We will need also the two-boundary of $\bar{\cal CD}(f,e)_{(3)}$.

\begin{itemize}

\item{}$\bar{\cal CD}(f,e)_{(2)}$ consists of triangles on the boundaries of tetrahedra in 
$\bar{\cal CD}(f,e)_{(3)}$ that link to, or are also in the boundary of,  tetrahedra outside of ${\cal CD}(f,e)$. 

\end{itemize} 

\subsection{Relations amongst causal diamonds}

Consider four events causally related as in
\f
d < e < f < g .
\ff

Then consider two causal diamonds, ${\cal CD}(d,g)$  and ${\cal CD}(f,e)$.

\begin{itemize}

\item{}We have,
\f
{\cal CD}(f,e) \subset {\cal CD}(g,d)
\ff

\item{}One can also show that there  exists a cross section $\Sigma_{g,d}$ of ${\cal CD}(d,g)$ and a cross section $\Sigma_{f,e}$ of 
${\cal CD}(f,e)$ such that
\f
\Sigma_{f,e} \subset \Sigma_{g,d}
\ff
Given a state $\rho (g,d) \in {\cal H}(g,d)$ we write the partial trace, 
\f
\rho (f,e) = Tr^{\prime} \rho (g,d)   \in {\cal H}(f,e)
\ff


\end{itemize}

\subsection{Elementary causal diamonds in spin foams}

An elementary causal diamond consists of two sequential events, $e < f$ or $ e \rightarrow f$ such that
\f
{\cal F S} (e) = {\cal P S} (f).   
\ff
There are then three time slices,
\f
\Sigma_1 = {\cal P S }(e), \ \ \ \ \ \Sigma_2 = {\cal FS} (e) = {\cal PS} (f), \ \ \ \ \Sigma_3 = {\cal F S} (f)
\ff

They have a common spatial boundary $W(f,e)$ which is a set of triangles.  Hence the area of the boundary is fixed:
\f
A(f,e) =\sum_{\triangle \in {\cal W} (f,e)} A_\triangle = \sum_{\triangle \in {\cal W} (f,e)} 8 \pi G \hbar \gamma {j_\triangle }
\ff
The volume of the elementary causal diamond is the volume of the middle slice, which is a sum of intertwiners on tetrahedra.
\f
V(f,e)= V ( \Sigma_2 ) = \sum_{T \in \Sigma_2 } (\hbar G)^{\frac{3}{2}} \hat{v}_T . 
\ff
where $\hat{v}_T$ is an operator in the space of intertwiners associated with each tetrahedra.

Examples are pairs of complementary Pachner moves:
\f
3 \rightarrow 2 \rightarrow 3, \ \ \  \ 2 \rightarrow 3 \rightarrow 2,  \ \ \ \ 1 \rightarrow 4 \rightarrow 1,  \ \ \ \ 4 \rightarrow 1 \rightarrow 4
\ff
Indeed, these are all there are, because of the structure of the Pachner moves.

Associated with the three slices are three finite dimensional Hilbert spaces, ${\cal H}_{I}$, $I=1,2,3$, in each of which there is generally a mixed state
$\rho_I \in {\cal H}_{I}$  (because the slice is an open system).  These Hilbert spaces are made from the dual spin networks, with the boundary spins fixed.

\subsection{Flat spin foams}

Below we will need to work with a spin foam history corresponding to flat a spacetime.  Fortunately there is available a characterization of a flat spin foam history, which makes use of the connection between Wieland's spin foam model\cite{WW3+1} and energetic causal sets\cite{ECS1,ECS2,ECS3}.  This formulation makes use of a future pointing normal $p_a^T$ associated to each tetrahedron, $T$ which is subject to the two constraints, (\ref{Pcon}) and (\ref{Ccon}).

The action for a Wieland spin foam has a part made of these constraints.
\f
{\cal S}=  [\sum_S z^a_S {\cal P}^S_a  
+\sum_T ( N_T {\cal C}_T + r_a^T \sum_{\triangle \in T} n^a_\triangle     )  + \sum_\triangle w_\triangle n_\triangle^a p_a^{T(\triangle)}]  
\ff
where $z^a_S$ are lagrange multilpliers.  They give an embedding of the events dual to the four simplicies in a four dimensional flat spacetime with metric $\eta_{ab}$ which is dual to the space the normals $p_a^T$ live in.  This embedding is determined by the equations of motion for the $p_a^T$ gotten from the variation of $\cal S$.
\f
\frac{\delta {\cal S}}{\delta p_a^T}= 0 \ \ \rightarrow \ \  z^a_{T^+} - z^a_{T^- } = N_T \eta^{ab} p_b^T + 
\sum_{\triangle \in T} w_\triangle n_\triangle^a 
\label{peom}
\ff
where $T^\pm$ are the two four simplices that share $T$, to its past and future, respectively.  Note that these depend on the $N_T$ which are also lagrange multipliers.

A simple counting argument suggests that the equations can generically always be solved to yield a spin foam history embedded in a flat spacetime.  Altogether there are $4 n_S + 5 n_T$ variables, the $p_a^T, N_T$ and $z^a_S$.  But $n_T = 2 n_S$ yielding $14 n_S $ variables.  (The volumes $V$ and potentials, $U$  are fixed functiond of the spins and intertwiners which are not counted here.)  These are exactly enough to solve $14 n_S$ equations, given by (\ref{Pcon}, \ref{Ccon}, \ref{peom}).  

In order that the mapping preserve the causal structure of the spin foam, the offsets, $\sum_{\triangle \in T} w_\triangle n_\triangle^a $ in (\ref{peom}) must leave the intervals $ z^a_{T^+} - z^a_{T^- } $  timelike and future pointing, which they otherwise are as the momenta
$p_a^T$ are both.   This can be achieved by setting the lagrange multipliers $w_\triangle =0$.  

Once the imbedding of the spin network into $M^4$ is accomplished the next task is to choose spins and intertwiners corresponding to a triangulation of $M^4$ matching the causal structure.

 

\section{Quantum gravity on a causal diamond}

We define the quantum theory on a causal diamond, defined by holding the boundary $\cal B$ fixed and summing over the degrees of freedom on the interior.  We define first the Hamiltonian theory; first the kinematical, then the physical Hilbert spaces.  Then we define the spin foam amplitudes that compute the physical expectation values.   

\subsection{The causal diamond Hilbert space and the boost energy}

Consider the waist of a causal diamond, which is a space like two surface, $W$.   
The waist bonds a family of space like three surfaces, $\Sigma $, such that $\partial \Sigma = W$, which span the interior of the causal diamond.  

In the spin foam we can decompose the two surface as a sum of triangles
\f
W= \sum_\tau  \triangle_\tau
\ff
while $\Sigma$ is composed of tetrahedra.

On each triangle there sits a representation of the lorentz group, ${\cal V}^{(\gamma (j+1),j)}$ and a unit normal in an internal flat space $n_a \in M^4$.

The kinematical Hilbert space of a causal diamond can be written as
\f
{\cal H}_{CD}^{kin} = \otimes_{\triangle \in W(CD)} {\cal V}_\triangle^j
 \otimes {\cal H}_{bulk}
\ff
where $ {\cal V}_{\triangle^j}$ is the simple representation $(\gamma (j+1), j)$ and the bulk portion ${\cal H}_{bulk}$ is made up of spin networks in the interior of a space like slice, $\Sigma$, with the topology of a disk, whose boundary matches $W$.

Loop quantum gravity tells us how to define operators on the kinematical Hilbert space, ${\cal H}_{CD}^{kin}$.  These include the quantum Hamiltonian and diffeomorphism constraints, ${\cal C}(N)$ and ${\cal D}(v)$ where $N$ is a density and $v^a$ is a vector field on $\Sigma$, both of which vanish on $W$.  
\f
N|_W = v^a|_W =0
\label{vanish1}
\ff

The physical Hilbert space, ${\cal H}_{CD}^{phys} \subset {\cal H}_{CD}^{kin}$
consists of states $|\Psi >$ in the kernel of these constraints,
\f
{\cal C}(N)|\Psi > = {\cal D}(v) |\Psi > =0, \ \ \ \ |\Psi > \in {\cal H}_{CD}^{phys}
\ff
We also impose the simplicity constraints on  all triangles, including those in the boundary, 
\f
<\Psi | {\cal S}^a |\Psi > =0, \ \ \ \ |\Psi > \in {\cal H}_{CD}^{phys}
\ff
where the inner product in ${\cal V}^{(\rho ,j)}$ is the usual one in which $\hat{L}^a$ and $\hat{L}^a$ are hermitian.

Now a lesson we learn from  FGP\cite{FGP} and Bianchi\cite{Eugenio1}, is that the right Hamiltonian to evolve a system in a near horizon region or, more generally, a causal diamond, is the generator of boosts in the boundary set, $W$ plus constraints acting in the interior.
\f
H_W^{boost}= \hbar \sum_{\triangle \in W} \hat{K}^a_\tau  n_a  + \mbox{constraints}
\label{boostH}
\ff
where $\hat{K}^a$ is a generator of boosts in $ {\cal V}_\triangle^j$ and $n_a$ is a unit normal in the internal space 
(with respect to the fixed internal lorentz metric).   

Note that because a boost is parametrized by a hyperbolic angle, the boost Hamiltonian has units of angular momentum.

In more detail we write
\f
H_W^{boost} (N,v) = \hbar \sum_{\triangle \in W} \hat{K}^a_\tau  n_a  + {\cal H}(N) +{\cal D}(v)
\label{boostH2}
\ff

The constraints act on  the space-like three surface that $W$ bounds.  These terms vanish when acting on physical states so that acting on physical states, the Hamiltonian that generates the generalized boost is a sum of boundary contributions. 

Physical observables on  ${\cal H}_{CD}^{phys}$, denoted ${\cal O}$, satisfy
\f
[{\cal O}, {\cal H}(N) ] = [ {\cal O}, {\cal D}(v) ]=0
\ff
Note that $\hat{A}(W) $ is physical, as is $H_W^{boost} (N,v) $ for all $N$ and $v^a$, that vanish on $W$. 

Among the physical observables are the area of the waist, $A(W)$.  One version of the area operator, appropriate for a spin foam model, is\cite{reviewSF}
\f
\hat{A}= 8 \pi \gamma G \hbar \sum_I \hat{L}_I^a n_a
\label{area}
\ff
expressed in terms of the operators for components angular momentum,  
$\hat{L}^a n_a$.   Here $\gamma$ is the Immirzi parameter.

\subsection{Generalized boosts}

Stripped of reference to a specific set of rigid coordinates here is what a boost is: {\it  A generalized boost, $B$ is a transformation that takes  a portion of a space like slice, $\Sigma_1$ to another portion of a space like slice, $\Sigma_2$ that share a common boundary,
$W = \partial \Sigma_1 =  \partial \Sigma_2$. } The resulting transformation between states is a boost operator.
\f
\hat{U}_B : | \Psi_1 > = |\Psi_2 >
\ff
where $\hat{U}_B (\eta ) = e^{-\frac{\imath}{\hbar}H_B (W) \eta }$
where $\eta$ is a dimensionless hyperbolic angle.

The usual definition of a boost in Minkowski spacetime satisfies this, as does a change of slicing of a causal diamond.  A boost captures the idea of bubble evolution, of  a local refoliation which affects a compact region of space,  $\Sigma_1$ and leaves the rest of space, $\Sigma_0'$ untouched.

In the case of causal diamonds, the transformations generated by $H_B(W)$ given by (\ref{boostH2}) are generalized boosts.

In a spin foam model a generalized boost which fixes a triangle, $\triangle$, is generated by a series of $1 \rightarrow 4$ moves which fix that triangle.  This is illustrated in Figures 3, taken from \cite{LS-thermobh1}, and Figure 4,  in the $1+1$ dimensional case, where a series of $1 \rightarrow 2$ moves fixes a dual vertex.  

The  $3+1$ boost move is illustrated in Figure 5 and 6.

By combining boost moves for triangles in $W$ we  can define spin foam histories that boost the interior of a  causal diamond, fixing its waist.  

\subsection{Topological field theory and simplicity relations}

The physical states are defined by summing over spin foam histories.  
In loop quantum gravity we describe the dynamics in terms of a sum over histories we call spin foams. Depending on the context and boundary conditions, we use a sum over spin foam histories to define a projection operator onto physical states, or physical evolution amplitudes.   

The basic idea of a spin foam model is to represent general 
relativity as a constrained topological field theory\cite{bc,FK,EPRL,reviewSF}.  
This holds both classically and quantum mechanically.  The constraints reduce the gauge freedom of a topological quantum field theory precisely in the right way as to induce the  two degrees of freedom per point of general relativity.  Given a history, expressed as a  four dimensional simplicial complex, $X$, labeled as above, we introduce the key features of the partition function
(below we add additional structure to write the full amplitudes, for the present we stick to the following simple form.)  The partition function for the topological $BF$ theory is
\f
{\cal Z}^{BF} (X) = \prod_{\triangle} \int d \rho_\triangle \sum_{j_\triangle} \prod_T \sum_{i_T} \prod_{S} {\cal A}_S (\rho, j, i )
\ff
where the amplitude ${\cal A}_S (\rho, j, i )$ attached to each four-simplex is a $10-j$ symbol, which satisfies recursion relations needed to make the partition function triangulation independent.  

The expectation value of a functional of the representations and intertwiners, ${\cal F}[\rho, j, i]$ can be expressed as
\f
< {\cal F} > =   \frac{1}{\cal Z}\prod_{\triangle} \int d \rho_\triangle \sum_{j_\triangle} \prod_T \sum_{i_T} \prod_{S} {\cal A}_S (\rho, j, i ) {\cal F}[\rho, j, i]
\ff

We then can write the simplicity constraint as follows.  For  every triangle, $\triangle$ of a history we impose
\f
0 = <{\cal S}^a_\triangle  >  
\label{simp2}
\ff
This defines a new partition function, which is taken to be a definition of quantum general relativity.  This is because we have implemented, in the measure of the state sum model for a $BF$ theory, the simplicity constraint that reduces that topological field theory to general relativity.
\f
{\cal Z}^{GR} (X) = \prod_{\triangle} \int d \rho_\triangle \sum_{j_\triangle} \prod_T \sum_{i_T} \prod_{S} {\cal A}_S (\rho, j, i ) \delta [\rho - \gamma ( j+1)  ] 
\ff

In the literature there are several different ways to impose the simplicity constraints in a spin foam model.  But they agree that the 
expectation value of the constraints vanishes, as in (\ref{simp2}).  
This expectation value is all we will need to derive the first law of thermodynamics below.  

\subsection{The full spin foam path integral}

We combine the constrained topological field theory structure with the Wieland causal structure.

The full amplitude for a given history is given by
\begin{eqnarray}
{\cal Z}^{GR} (X) &=&  \prod_{\triangle} \int d \rho_\triangle \sum_{j_\triangle} dn_\triangle^a
\delta ({\cal S}_\triangle ) 
 \delta (n^a n^b \eta_{ab} -1 ) dw_\triangle  {\cal A}_\triangle (\rho, j, i ) \ 
\nonumber \\
&& \prod_{T = tetrahedra}  \sum_{i_T} \int dp^T_a \int dN_T dr_T  {\cal A}_T (\rho, j, i ) \ 
\nonumber \\
&& \prod_{S=four-simplex} \int dz_S^a  
 {\cal A}_S (\rho, j, i ) \ 
 \ \nonumber \\
&& \times 
 e^{\imath [\sum_S z^a_S {\cal P}^S_a  
+\sum_T ( N_T {\cal C}_T + r_a^T \sum_{\triangle \in T} n^a_\triangle     )  + \sum_\triangle w_\triangle n_\triangle^a p_a^{T(\triangle)}]   }
\label{fullZ}
\end{eqnarray}

The general model is made precise by specifying the amplitudes, $ {\cal A}_S (\rho, j, i )$, ${\cal A}_T (\rho, j, i )$ and 
$ {\cal A}_\triangle (\rho, j, i ) $.  The preferred  choices for these are discussed in \cite{reviewSF}.  These change in any case under renormalization, as we describe now briefly.  But the important point is that 
the results concerning the second law we describe in the next section are to a large extent independent of these amplitudes,
as long as the simplicity constraints are imposed and there is a semiclassical limit.


\subsection{Observables on the boundary of a causal diamond}

We can define a set of observables of quantum gravity, expressed in terms of a causal diamond ${\cal C D}$.   

We first define a causal diamond spin foam to be a causal spin foam history, $X$ which has the structure of a
causal diamond.  We fix a  particular $X$ and extract its two boundary $\bar{X}_{(2)}= \bar{\cal CD}(f,e)_{(2)}={\cal B}$.
Now we fix a particular $\cal B$ and consider the ensemble of spin foam histories ${\cal X}_{\cal B}$ to
consist of all $X$ such that $\bar{X}_{(2)} ={\cal B}$.

${\cal B}$ splits into three components as described above
\f
{\cal B} = {\cal I}^+ \cup {\cal I }^- \cup \bar{\cal W}
\ff

On each triangle $\triangle \in {\cal B}$ we fix a normal $n_\triangle$.

Consider a set of currents
on the past and future null boundaries, ${\cal J}^-, {\cal J}^+$ and ${\cal J}^W$ on the spatial boundary $W$.   
We will assume that these couple to areas, but there are also other possibilities. The partition function for 
a fixed spinoff history, $X$, is 

\begin{eqnarray}
{\cal Z}^{GR} [X, {\cal J}]  &=&  \sum_{X \in {\cal X}_{\cal B}} \prod_{\triangle} \int d \rho_\triangle \sum_{j_\triangle} 
[dn_\triangle^a ]^\prime
\delta ({\cal S}_\triangle ) 
 \delta (n^a n^b \eta_{ab} -1 ) dw_\triangle  {\cal A}_\triangle (\rho, j, i ) \ e^{\imath \sum_{\triangle \in {\cal B}} {\cal J}_\triangle j_\triangle }
\nonumber \\
&& \prod_{T = tetrahedra}  \sum_{i_T} \int dp^T_a \int dN_T dr_T  {\cal A}_T (\rho, j, i ) \ 
\nonumber \\
&& \prod_{S=four-simplex} \int dz_S^a  
 {\cal A}_S (\rho, j, i ) \ 
 \ \nonumber \\
&& \times 
 e^{\imath [\sum_S z^a_S {\cal P}^S_a  
+\sum_T ( N_T {\cal C}_T + r_a^T \sum_{\triangle \in T} n^a_\triangle     )  + \sum_\triangle w_\triangle n_\triangle^a p_a^{T(\triangle)}]   }
\label{fullZJ}
\end{eqnarray}

Here $[dn_\triangle^a ]^\prime$ means we do not integrate over the normals of triangles in the boundary, as those are fixed.

To get the full amplitude we sum over all causal spin foam histories with the same boundary
\f
{\cal Z}^{GR} [{\cal B}, {\cal J}]   = \sum_{X \in {\cal X}_{\cal B}} {\cal Z}^{GR} [X, {\cal J}] 
\ff

${\cal Z}^{GR} [{\cal B}, {\cal J}] $ is a generating functional for scattering of gravitational degrees of freedom across the causal diamond.

We define the vacuum expectation value of a function of the boundary observables as

\begin{eqnarray}
< {\cal F}( \rho_\triangle, j_\triangle ) > &=&  \sum_{X \in {\cal X}_{\cal B}} \prod_{\triangle} \int d \rho_\triangle \sum_{j_\triangle} 
[dn_\triangle^a ]^\prime
\delta ({\cal S}_\triangle ) 
 \delta (n^a n^b \eta_{ab} -1 ) dw_\triangle  {\cal A}_\triangle (\rho, j, i ) 
 \\ \nonumber
&&  {\cal F}( \rho_\triangle, j_\triangle )  e^{\imath \sum_{\triangle \in {\cal B}} J_\triangle j_\triangle }
\nonumber \\
&& \prod_{T = tetrahedra}  \sum_{i_T} \int dp^T_a \int dN_T dr_T  {\cal A}_T (\rho, j, i ) \ 
\nonumber \\
&& \prod_{S=four-simplex} \int dz_S^a  
 {\cal A}_S (\rho, j, i ) \ 
 \ \nonumber \\
&& \times \left ( 
 e^{\imath [\sum_S z^a_S {\cal P}^S_a  
+\sum_T ( N_T {\cal C}_T + r_a^T \sum_{\triangle \in T} n^a_\triangle     )  + \sum_\triangle w_\triangle n_\triangle^a p_a^{T(\triangle)}]   } \right ) _{{\cal J}=0 }
\label{fullZJ5}
\end{eqnarray}

\subsection{Renormalization}

We now sketch the processes of renormalization and coarse graining, as mentioned in the introduction, we do this mainly to show that the first law for  causal diamonds emerges in a way which is largely independent of those processes.  

Given a given spin foam history, $X$ we may act with a five dimensional Pachner move, ${\cal P}^5$ to yield
another spin foam history with the same topology and boundary, $X^\prime = {\cal P} \circ X$. 
The equivalent amplitude for $X^\prime$ is 
\begin{eqnarray}
{\cal Z}^{GR \prime} [X^\prime, {\cal J}]  &=&  \sum_{X \in {\cal X}_{\cal B}} \prod_{\triangle} \int d \rho_\triangle \sum_{j_\triangle} [dn_\triangle^a ]^\prime
\delta ({\cal S}_\triangle ) 
 \delta (n^a n^b \eta_{ab} -1 ) dw_\triangle  {\cal A}_\triangle (\rho, j, i )^\prime
  \ e^{\imath \sum_{\triangle \in {\cal B}} J_\triangle j_\triangle }
\nonumber \\
&& \prod_{T = tetrahedra}  \sum_{i_T} \int dp^T_a \int dN_T dr_T  {\cal A}_T^\prime (\rho, j, i ) \ 
\nonumber \\
&& \prod_{S=four-simplex} \int dz_S^a  
 {\cal A}_S (\rho, j, i )^\prime \ 
 \ \nonumber \\
&& \times 
 e^{\imath [\sum_S z^a_S {\cal P}^S_a  
+\sum_T ( N_T {\cal C}_T + r_a^T \sum_{\triangle \in T} n^a_\triangle     )  + \sum_\triangle w_\triangle n_\triangle^a p_a^{T(\triangle)}]   }
\label{fullZJ2}
\end{eqnarray}

Note that as gravity is not a topological field theory, ${\cal Z}^{GR \prime} [X^\prime, {\cal J}] $ is not equal to the 
original amplitude evaluated on $X^\prime$, 
\f
{\cal Z}^{GR \prime} [X^\prime, {\cal J}]  \neq {\cal Z}^{GR } [X^\prime, {\cal J}] 
\ff

Pachner's 
theorem tells us that given any two spin foam histories, $X_1$ and $X_2$ with the same topology and boundary there
is a sequence of Pachner moves, ${\cal P}_I$ that connects them. Thus, beginning with a definition of the fundamental theory in terms of an vertex amplitude, one arrives at a coarse grained renormalized amplitude.

This is defined, as emphasized in [\cite{Bianca1}], in terms of a coarse graining of the boundary.  So we now focus on how to do that.

\subsection{Coarse graining of the spatial boundary}

We have a simple point to make, which is the invariance of the linear simplicity constraints on the boundary, under coarse graining of the boundary.

Let us consider a grouping of the triangles of $\bar{\cal W}$ into a fewer number of larger triangles, $\tau^\prime$.
We  have
\f
\triangle^{\prime}_I = \cup_{\triangle_i \in \{ \triangle^\prime_I \}} \triangle_i
\ff
We define this to be the coarse grained waist. Let ${\cal B}^\prime$ be a course graining of the boundary
$\cal B$ that contains this  coarse graining of the waist.  Given a spin foam history $X$ with boundary
$\cal B$, there is a coarse graining of $X$, labeled $X^\prime$ that has the boundary ${\cal B}^\prime$, which
can be reached by a sequence of Pachner moves.  

We define the operators on the new coarse grained triangles to be the sums of those of their constituents,
\f
\hat{L}^a_{\triangle^\prime} = \sum_{\triangle \in \triangle^\prime} \hat{L}^a_{\triangle}, \ \ \ 
\hat{K}^a_{\triangle^\prime} = \sum_{\triangle \in \triangle^\prime} \hat{K}^a_{\triangle},
\ff

We note that for any such coarse graining the simplicity constraint of any boundary triangle is respected
\f
\boxed{
< {\cal S}^a_{\triangle^\prime_I }> = 0 }
\label{simp7}
\ff

This is a key, if simple, result, as it means that the connections with the first law of thermodynamics we are about to discuss are stable under renormalization and coarse graining of the boundary.

Let us now presume that a particular state yields an effective geometry which has a radius of curvature, $R$, large 
in Planck units, $R >> l_P$.  Then we can expect to be able to coarse grain up to a scale,
$L$ satisfying,   $l_P << L << R$.   In this case we can expect to be able to choose the normals on the boundary 
triangles so that $| \nabla_a n^b |  < \frac{1}{R} $.  

This further implies that $| \nabla_a n^b |  < \frac{1}{l} $ where $l$ is the scale of the causal diamond itself, since we require that
$l << R$.  So if the coarse grained triangles are large on the Planck scale, but small compared to $l$ and $R$, we  can ignore the variations in the normals of the traingles making up the coarse grained triangles.

We then may choose the normals in each of the larger triangles to be equal, up to terms of order $\frac{1}{R}$.
\f
\{ i,j \} \in I \rightarrow n^a_{\triangle_i } = n^a_{\triangle_j } = n^a_{\triangle_I^\prime }
\label{equal}
\ff

We then have also, up to terms of order $\frac{1}{R}$,
\f
\hat{L}^a_{\triangle^\prime} n_a^{\triangle^\prime}= \sum_{\triangle \in \triangle^\prime} \hat{L}^a_{\triangle}  n_a^{\triangle^\prime}. 
\ff

\section{The simplicity constraint and  the first law}

Using the result above, we can establish several  relationships between the simplicity constraint (\ref{simp}) and the first law.  

Stripped of the technical details, the point is very simple: {\it the simplicity constraint of spin foam models is equivalent to the first law of thermodynamics, in the sense that they imply each other.}  Indeed this was implicit in Bianchi's derivation of horizon entropy\cite{Eugenio1}, as well 
as in the Frodden-Gosh-Perez papers\cite{FGP}.

\subsection{Derivation of the first law of quantum spacetime from the simplicity constraint}

We begin with the simplicity constraint, which can be expressed as follows.  For every space like triangle in a spin foam history,
we have the simplicity constraint,
\f
 < \hat{\cal S}^a_\triangle > =0 
 \label{simp4}
\ff

We consider triangles in the waist.  Each has a fixed unit normal, $n_{\triangle a }$, we multiply by these
\f
 < \hat{\cal S}^a_\triangle n_{\triangle a }> =0 
 \label{simp5}
\ff

Sum this over the triangles in the wast of a causal diamond,, $W=\sum_\triangle \triangle $ 
\f
 <\hat{\cal S}(W) >= < \sum_\triangle \hat{\cal S}^a_{\triangle}  n_{\triangle a }> 
\ff
This means
\f
 < \sum_\triangle \hat{K}^a_{\triangle}  n_{\triangle a }>= \gamma  < \sum_\tau \hat{L}^a_{\triangle}  n_{\triangle a }>
\ff
Multiply by $\hbar$, use the definition of the boost Hamiltonian (\ref{boostH2}) and the 
area of $W$ (\ref{area}) to find that on physical states,
\f
\boxed{< H_B (W)  > = \left ( \frac{\hbar}{2\pi } \right ) \frac{<\hat{A}(W)>}{4 G \hbar}}
\label{firstly}
\ff
We call this the  {\it  first law of quantum spacetime.}  It is the quantum version of the first law of classical 
spacetime (\ref{boost1}).

Note also that (\ref{simp7}) holds for any coarse graining of the boundary, and after an arbitrary number of renormalization steps.
We sketched the coarse graining and renormalization processes just to indicate that the result that the first law in the form
of (\ref{firstly}) holds as a consequence of the simplicity constraint is independent of those processes.  In particular, the first law holds at the level of coarse grained, renormalized observables.  This is because it is a consequence of a constraint which is linear in terms of both bare and renormalized quantities.  

In particular, this is due to the assumption that the normals of the triangles, $n^a_\triangle$ can be taken to be constant over the 
coarse grained boundary triangles.  There will be higher order corrections coming from terms in $\partial_b n^a_\triangle$.

\subsection{The microcanonical entropy}

To go further we must distinguish different ensembles.  As the area of the waist is a physical observable, we can define
the ensemble at fixed area.  We may call this the micro canonical ensemble.  

The boundary Hilbert space on $W$ is finite dimensional once the simplicity constraints have been imposed, since it is 
equivalent to the $LQG$ boundary Hilbert space.  The microcanonical entropy is defined as the log of the dimension of
the Hilbert space.  This has been computed to be proportional to the area, 
\f
S_{micro} (W) = \frac{<\hat{A}(W)>}{4 \alpha G \hbar}
\ff
where $\alpha$ depends on the Immirzi parameter.  Different assumptions lead to slightly different  values of $\alpha$,
all of order unity.  

Once we postulate this microcanical
the quantity in parentheses then has to be identified as a boost temperature in the microcanical ensemble at fixed area.
\f
T_{micro} = \frac{\hbar\alpha }{2\pi }
\ff

We stress  that the argument we have given yields (\ref{firstly}).  
Thus, this argument fixes the product $T_U \delta S$.  If either of these has an independent derivation, the other one is fixed.

But there is an independent derivation of the temperature, given by \cite{Eugenio1}, to which we now turn.

\subsection{The temperature of a causal diamond:}

Instead of working with fixed area, we can work with the canonical ensemble at fixed temperature.
To do this, we need an independent computation of the temperature.  To get this, we can note
that
the description of a quantum causal diamond we have given is ver similar to  Bianchi's near horizon quantum Rindler spacetime\cite{Eugenio1}.  
There he follows the path of Unruh and deWitt and couples a two state detector to the boundary state and computes the temperature of the detector in equilibrium.  This computation is done by computing the transition amplitude for exciting the detector, as a function of time, as measured by a clock carried by the detector.

We can then consider exactly the same process, and couple the boundary state of the causal diamond to a two state detector and compute the temperature of the detector.  As the Hilbert of the boundary triangles of the causal diamond are the same as considered by Bianchi, we can
apply his result, computed in \cite{Eugenio1}, equations 10 to 18.  This leads to the conclusion 
that the state is hot, with an (angular)  Unruh temperature.
\f
T_U  = \frac{\hbar}{2 \pi c}
\label{aUnruh}
\ff
We note that because all boosts are equivalent in this context, there is just a single, angular, Unruh-like temperature.
There is no refoliation invariant meaning that could be given to the acceleration of an observer, usually denoted $a$.
Hence we can give no meaning to an Unruh temperature,  if by that we mean a quantity with units of energy.  But we can give
a meaning to an angular temperature like (\ref{aUnruh}), with units of angular momentum.   This makes sense because a boost translates in a hyperbolic angle, which is dimensionless.

The result is that the entropy must be identified as 
\f
S_{canonical} (W) = \frac{<\hat{A}(W)>}{4 G \hbar}
\ff

We then have from (\ref{firstly}) the {\it canonical first law of quantum spacetime.}
\f
\boxed{< H_B (W)  > = T_U 
S_{canonical} (W) }
\label{firstly2}
\ff



\subsection{Thermalization and the Boltzman entropy}

We next  investigate the thermallization of the state associated with a  causal diamond.   

Now let us consider density operators, $\rho$  on  ${\cal H}_{CD}^{phys}$.  These satisfy
\f
[ \rho , {\cal H}(N) ] = [ \rho , {\cal D}(v) ]=0
\ff
as well as 
\f
[ \rho , {\cal S}^a ]=0 .
\label{sim17}
\ff

We are interested in states that describe equilibrium, these are states that the system reaches after arbitrary boosts.  These are analogous to the quantum Rindler states of \cite{Eugenio1}.  
Once a state has been boosted sufficiently it is not going to be changed by boosting it further.  Thus, these should satisfy
\f
[ \rho_E , H_W^{boost} (N,v) ] =0
\ff
for all $N$ and $v^a$ which satisfy (\ref{vanish1}).   Such states are given by
\f
\rho_E (N,v) =   e^{-\frac{2 \pi}{\hbar} H_W^{boost} (N,v)}
\label{rhoE}
\ff
Note that the temperature, $\beta^{-1} = T_U = \frac{\hbar}{2\pi}$ is determined independently by comparison with Bianchi's calculation in (\cite{Eugenio1}).  Once this coefficient is determined there remains the freedom of choosing $N$ and $v^a$, subject to the condition that they vanish on $W$.  However, note that,
so far as physical observables are concerned, these are all equivalent to each other, because we can use the quantum constraints to transform them into each other.  We have
\f
Tr [ {\cal O} \rho_E (N,v) ] = Tr [ {\cal O} \rho_E (N',v') ].
\ff
Thus,  because of the many fingered time invariance, there is only one physically distinct boost generator.   We can define
\f
<{\cal O} >_{boost} =Tr [ {\cal O} \rho_E (N,v) ]
\ff
using any smooth $N$ and $v^a$ that vanishes on $W$.   

This boosted state behaves like a Rindler state.  Notice that because the Hamiltonian is modular,we can compute the Boltzmann statistical
mechanical 
entropy directly
\f
S(W)_{stat} = - Tr \rho_E \ln \rho_E = \frac{< H_W^{boost} > }{T_U}
\ff
This is the genuine first law of statistical thermodynamics.
If we put this together with the previous results we can deduce that
\f
S(W)_{stat} =\frac{<\hat{A}(W)>}{4 G \hbar}
\label{Sboltz}
\ff



\subsection{The thermodynamic first law}

We now finally are in a position to present an argument that the first law of thermodynamics itself, as a principle of statistical thermodynamics, is a consequence of the simplicity constraints, plus some natural assumptions.  

Let us compare two quantum states of a causal diamond.  

1) The vacuum, which we assume is dominated by flat spin foam history, constructed as described in section (3.7) with no matter, and 

2) a low
lying excitation of the flat spin foam whose gravitational field is weak.  This means that the quantum state of the c causal diamond is in the semiclassical regime, which is  defined as follows.   

Let $R (T) = < \frac{1}{\sqrt{{\cal R}(T)}} >$ be the curvature scale of a tetrahedron and similarly, $R (f,e) =<  \frac{1}{\sqrt{{\cal R}(f,e)}} > $ be the curvature scale of a causal diamond.

A spin foam history is in the semiclassical domain if there is a scale $l>> l_P$ such that for every causal diamond such that
\f
V(f,e) \approx l^3,
\ff
and the curvature scales of that causal diamond and all of its tetrahedra satisfy
\f
R(f,e) >> l >> l_P, \ \ \  R(T) >> l >> l_P
\label{semi}
\ff

We will assume that both states have the same volume.

We assume also that the quantum constraints are satisfied, as are the  classical constraints to leading order.  Given
(\ref{semi}) we can deduce that the contributions to the Hamiltonian constraint coming from gravitational radiation may be neglected.  That is, we assume that to leading order the boot energy is dominated by the heat  flow, so that
\f
\delta <H_B(W) > = <H_B(W) >_2 - <H_B(W) >_1 =  \delta Q = Q_2 -Q_1 
\ff
A calculation (\cite{Ted2015}) shows $\delta Q$ is given by an averaged energy-momentum tensor by
\f
\delta Q= - \frac{\Omega_{(2)}l^4}{15} <T_{ab}> t^a t^b 
\label{deltaQ}
\ff

is the energy from matter flowing through the causal diamond.  Here $t^a$ is the normal to a slice $\Sigma$.   

But by the first law of quantum spacetime,
\f
\boxed{
\delta <H_B(W) > = T_U \delta S_B }
\label{thermofirst}
\ff
where the change in the Bekenstein entropy is
\f
\delta S_B = S_2 - S_1
\ff

However, on the assumption that the state of the causal diamond is the equilibrium state (\ref{rhoE}), the Bekenstein entropy is the Boltzman entropy.  Thus we arrive at the standard first law of statistical thermodynamics, relating the heat flow, the temperature and the Boltzmann entropy.  Gravity no where appears in this relation, which indeed contains no $G$,  but its origin in this context is gravity, indeed we have traced it to the simplicity constraint of quantum gravity.

\subsection{Deriving the simplicity constraints from the first law of thermodynamics}

Now we go the reverse way, from the first law to the simplicity constraints.

Start with a classical solution to the Einstein equations, with arbitrary matter and pick a causal diamond, 
${\cal CD}(f,e)$, based on two causally related events, $f>e$.  It has been shown that the first law of thermodynamics holds on the waist, ${ W}(f,e)$ of ${\cal CD}(f,e)$,
\f
H_B(W) = T_U S (W)
\ff
Note that the $\hbar$'s cancel, so the classical relation is actually what might be called the first law of classical spacetime,
\f
H_B(W) = \frac{1}{8 \pi G} A (W)
\ff
Expressed in terms of a single spin foam history, in which $W(f,e)$ is a union of triangles,  $W =\cup \triangle$, this is,
\f
\sum_{\triangle \in W } \left ( H_B(\triangle ) - \frac{1}{8 \pi G} A ( \triangle ) \right ) =0 .
\label{first7}
\ff
But the quantum theory is defined by a sum over histories.
Moving (\ref{first7}) inside the path integral express this as an expectation value.
\f
<  \sum_{\triangle \in W } \left ( \hat{H}_B(\triangle ) - \frac{1}{8 \pi G} \hat{A} ( \triangle ) \right ) > =0 .
\ff
Since the path integral defines a projection operator on physical states, we may assume that the state is physical.  Hence the boost Hamiltonian is represented by its boundary term.  Meanwhile, we express the area in terms of the area operator.  Factoring out an $\hbar$, this gives us
\f
<  \sum_{\triangle \in W } \left ( \hat{K}^a _\triangle n_{a \triangle} -\gamma \hat{L}^a _\triangle n_{a \triangle}\right ) > =0 .
\label{sum1}
\ff
But any single triangle could be part of many waists of causal diamonds, each with different normals.  Hence, (\ref{sum1}) has to hold for each triangle and normal and we have derived the simplicity constraints
\f
< \hat{\cal S}^a > = < \left ( \hat{K}^a _\triangle  -\gamma \hat{L}^a _\triangle \right ) > =0 
\ff

\subsection{The principle of maximal entanglement}

We  can understand the unique equilibrium state, (\ref{rhoE}) and unique Unruh temperature, (\ref{aUnruh}),  in the following way.

Consider a causal diamond, ${\cal C}(f,e)$,  small compared to the radius of curvature, but large on the Planck scale.  We can consider it as embedded in any number of larger causal diamonds as described above.  Let's call $A = {\cal C}(f,e)$,
$C= {\cal C}(g,d)$ and $B= {\cal C}(g,d) - {\cal C}(f,e)$.   

Now we reason by analogy to flat spacetime.   Start with a generic $\rho_{(g,d)}$ on ${\cal H}(g,d)$.  
What happens when we make a boost, defined by the condition that we fix the two surface which is the waist of the
smaller causal diamond?   The result must be to define the state on the smaller diamond by tracing out the degrees of 
freedom external to it.   

Define
this reduced density matrix on $A$ as usual by
\f
\rho_A = Tr_B \rho_{A+B}.
\ff
The principle of maximal entanglement says that for large $C$ and small $A$ the state $\rho_A$ has a minimal amount of information in it as to physics in $A$.   i.e the state $\rho_{A+B}$ was maximally entangled, so when we trace by $B$ and destroy all the corelations generated by that entanglement we have no information left.  Thus, by analogy with the situation in Rindler spacetime, this must be a thermal state.   Since all boosts that fix $W$ but boost its interior are equivalent, 
we can conjecture that
\f
\rho_A = \rho_E =  e^{-H_{boost} (W) /T_U}.
\label{thermal17}
\ff
There is just a single  choice for this state because all boosts are equivalent, so up to gauge transformations all
boost Hamiltonians are equivalent 

We also see that if the global state is maximally entangled this gives rise to a universal boost temperature.
\f
T_U = \frac{\hbar}{2 \pi c}
\ff

\subsection{Proposal for  a quantum equivalence principle}

This last result can be reformatted as a statement of the quantum equivalence principle.

\begin{itemize}

\item{} Let us consider a pure physical state of the quantum gravitational field $\rho$ holding in a region, $\cal R$, of spacetime.
Let $A= {\cal CD}(f,g)$ be a causal diamond within that region, with waist, $W$, defined by two events, $f$ and $e$.  Let
$B$ be the complement of $A$ in $\cal R$.    Then the state 
\f
\rho_A = Tr_B \rho
\ff
is maximally entangled with degrees of freedom in the complement, $B$.

To see this, let $H_{boost} (W)$ be the quantum Hamiltonian that generates generalized boosts in $A$, leaving $W$ fixed.   Then 
(\ref{thermal17}) holds.

\end{itemize}

This is a version of the equivalence principle because it says that a general boosted observer in a quantum spacetime sees the same thing that a boosted observer in flat spacetime sees, namely that the region inside the surface fixed by the boost is maximally entangled with the region in the exterior of that surface.  

\section{The recovery of GR}

Our final step is to use the results gotten so far to  understand why the Einstein equations must characterize the semiclassical limit of spin foam  models.  We do not show that such a limit exists, but we do show that if it does,  its dynamics are captured by the semiclassical Einstein's equations.  We get to this result by following Jacobson in \cite{Ted2015}.

We work in the semiclassical regime described above in section (5.5).  We consider as in that section a comparison between a flat causal diamond and a low energy excitation which is describable in terms of classical fields, slowly varying on the  scale of the causal diamond.  This low energy excitation has the same volume, but a different area.  The variation of the area at fixed volume can be related to the spatial scalar curvature averaged on the causal diamond\cite{GaryG,Ted2015}. 
\f
\delta A (f,e) = A(f,e) - A_{flat} (V) = - \frac{\Omega_{(2)}l^4}{30}  {\cal  R }(f,e)
\ff
We follow Jacobson\cite{Ted2015} in describing the geometry of the causal diamond in Riemann normal coordinates.  In that case the extrinsic curvature can be taken to vanish to leading order and we have
\f
{\cal  R }(f,e) = 2 G_{ab} t^a t^b
\ff
where $G_{ab}$ is the Einstein tensor.   The heat flow is given by (\ref{deltaQ}).   Plugging these relations into the first law (\ref{thermofirst}) we find
\f
t^a t^b \left ( G_{ab} - 8 \pi G < T_{ab} >  \right ) =0 
\label{almosteinstein}
\ff 
Jacobson points out that the remaining steps of the derivation are simplest if we impose that the matter is conformally invariant\cite{Ted2015}.
\f
g^{ab} < T_{ab} >  =0
\ff
(We refer the reader to \cite{Ted2015} for the case of non-conformally invariant matter, as well as the case of non-vanishing cosmological constant.)   In this case we can argue that since
$R >> l$, within one curvature scale there will be many causal diamonds, with different normals $t^a$.  This means we can remove the $t^a$ to find
\f
 G_{ab} = 8 \pi G < T_{ab} > 
 \ff

\section{Conclusions}

By extending results of \cite{FGP,Eugenio1,carloetal}, 
we have shown that, given suitable conditions, the linear simplicity constraint of spin foam models, (\ref{simp}) implies the first law of quantum spacetime (\ref{firstly}).  This is initially expressed in terms of a micro canonical entropy, given by the ensemble at fixed area.  But if we go to the canonical ensemble at fixed temperature we can follow \cite{Eugenio1} to compute the temperature. and from that, the entropy and assign the causal diamond to an equilibrium state at fixed temperature.  This then implies  the relationship between the area and the Boltzmann entropy (\ref{Sboltz}), with the correct $\frac{1}{4}$, independent of the Immirzi parameter.   

We have also showed that the first law of thermodynamics implies the simplicity constraint.  

We further showed that {\bf if} there exists a semiclassical limit (which we do not prove) this implies the thermodynamic first law, 
(\ref{thermofirst}).   This, in turn, implies the Einstein equations, as shown by Jacobson in \cite{Ted2015}.  

These results  establish that there is a close connection between the holographic behaviour of quantum gravity and the fact that general relativity is closely related to a topological field theory.  
Indeed, this is precisely the connection  anticipated in \cite{Louis1} and \cite{linking}, 
The fact that general relativity is a constrained topological field theory is then the root of the holographic nature of gravity.   Indeed, this  has been since then a central feature of loop quantum gravity\cite{linking}, which has been developed in different ways in  \cite{boundary}.  It is fitting that this connection between the holographic and topological aspects of gravity is deepened by the simplicity constraints, which were also first used in works of Barrett and Crane\cite{bc}.  

There is one big question that these results raise, which is that if general relativity, which is a time reversible theory, corresponds to equilibrium statistical mechanics, what is the time irreversible extension of general relativity that corresponds to non-equilibrium statistical mechanics\footnote{Nonequilibrium extensions of black hole thermodynamics are discussed in \cite{laurentnon}.}?  In particular, might it be one of the known irreversible extensions of general relativity\cite{ire}?

\section*{ACKNOWLEDGEMENTS}

It is a pleasure to thank Bianca Dittrich, Laurent Freidel, Ted Jacobson, Rob Myers, Alejandro Perez,  Aldo Riello, Carlo Rovelli and Wolfgang Wieland for comments and discussion.  

This research was supported in part by Perimeter Institute for Theoretical Physics. Research at Perimeter Institute is supported by the Government of Canada through Industry Canada and by the Province of Ontario through the Ministry of Research and Innovation. This research was also partly supported by grants from NSERC, FQXi and the John Templeton Foundation.

\end{document}